\documentstyle[preprint,aps]{revtex}

\tightenlines

\begin{document}
\title{Exact Solution of Photon Equation in Stationary G\"{o}del-type and G\"{o}del
Space-Times}
\author{Ali Havare, Taylan Yetkin}
\address{Department of Physics, Mersin University, 33342, Mersin-Turkey}
\date{\today }
\maketitle

\begin{abstract}
In this work the photon equation (massless Duffin-Kemmer-Petiau equation) is
written expilicitly for the general type of stationary G\"{o}del space-times
and is solved exactly \ for the G\"{o}del-type and the G\"{o}del
space-times. Harmonic oscillator behaviour of the solutions is discussed and
energy spectrum of photon is obtained.

\end{abstract}

\section{Introduction}

The relativistic particles satisfying wave equations in cosmological
backgrounds are considered to analyze quantum effects in curved space-times.
For this purpose single particle states are studied in detail and also some
restrictions on solutions are examined to obtain behavior of the particles.
In the atomic scale relativistic wave equations which conformed to general
relativity may not be important because of the weakness of gravitational
effects. On the contrary for many astrophysical situations one has to take
into account gravitational effects due to it's dominant role; for example
particle creation by black holes.

The most studied relativistic equations are Klein-Gordon and Dirac equations
describing spin zero and spin one-half particles respectively. These
equations are considered mostly in the expanding universes \cite{1,2,3,4,5,6}
which are members of Friedmann cosmological models. Other type of universes
are so-called G\"{o}del and G\"{o}del-type space-times where they have a
global rotation in all point of the space. Theoretical works of Gamow \cite
{7} and G\"{o}del \cite{8}, and some other evidences of rotation \cite
{9,10,11} are directed us to study the behavior of electromagnetic fields 
\cite{12,13,14,15} and spinning particles in curved rotating backgrounds.
The line-elements of the G\"{o}del space-time 
\begin{equation}
ds^{2}=-(dt+e^{ar}d\theta )^{2}+dr^{2}+\frac{1}{2}(e^{ar}d\theta
)^{2}+dz^{2},  \label{1}
\end{equation}
and the G\"{o}del-type space-time 
\begin{equation}
ds^{2}=-(dt+md\theta )^{2}+dr^{2}+(l+m^{2})d\theta ^{2}+dz^{2}  \label{2}
\end{equation}
are studied for Klein-Gordon and Weyl equations \cite{16,17,18,19}. The
functions $m$ and $l$ in the G\"{o}del-type line-element are only functions
of $r$ and they satisfy following two independent conditions \cite{15}: 
\begin{equation}
D=\left( l+m^{2}\right) ^{1/2}=A_{1}\exp (ar)+A_{2}\exp (-ar),  \label{3}
\end{equation}
\begin{equation}
\frac{1}{D}\frac{dm}{dr}=C  \label{4}
\end{equation}
or 
\begin{equation}
D=Ar,\frac{1}{D}\frac{dm}{dr}=C,  \label{5}
\end{equation}
where $A_{1},A_{2},A$ and $C$ are arbitrary constants. Second condition
gives the singular homogeneous G\"{o}del-type solution, but it can not be
obtained from the first condition. As a particular choice of the first
condition one takes $a=0,$ then 
\begin{equation}
D=const.  \label{6}
\end{equation}
If we choose $A_{1}=1/\sqrt{2}$ and $A_{2}=0,$ then $D=(1/\sqrt{2})\exp
(ar). $ This gives $m=(C/\sqrt{2}a)\exp (ar)$ (with a constant of
integration equal to zero) and if we take $C=-a\sqrt{2},$ then $m=\exp (ar)$
and $l=(-1/2)\exp (2ar),$ that gives G\"{o}del solution (with $C<0$)\cite{8}.

Although there are some works concerning with the electromagnetic fields in
these space-times there is not any attempt to find exact solution of the
quanta of electromagnetic field, i.e. photon.\ The photon is a massless
spin-one particle and it obeys massless case of Duffin-Kemmer-Petiau (DKP)
equation which can be formed as 
\begin{equation}
\left[ i\beta ^{(\alpha )}\partial _{(\alpha )}+m\right] \Psi =0  \label{7}
\end{equation}
where $\beta ^{(\alpha )}=\gamma ^{(\alpha )}\otimes $I$+$I$\otimes \gamma
^{(\alpha )}.$ This $16\times 16$ matrix equation is in Minkowski frame and
it must be conformed to general relativity to find out quantum effects in
curved backgrounds. In the studies of \"{U}nal \cite{20} and Lunardi et al. 
\cite{21} it has been shown that the covariant form of DKP equation is given
by 
\begin{equation}
(i\beta ^{\mu }\nabla _{\mu }+m)\Psi =0  \label{8}
\end{equation}
where $\beta ^{\mu }(x)=\gamma ^{\mu }(x)\otimes $I $+\,$I$\,\otimes \gamma
^{\mu }(x)$ are the Kemmer matrices in curved space-time and they are
related to flat Minkowski space-time as 
\begin{equation}
\beta ^{\mu }=e_{(\alpha )}^{\mu }\beta ^{(\alpha )}  \label{9}
\end{equation}
with a tetrad frame that satisfies 
\begin{equation}
g_{\mu \nu }=e_{\mu }^{(\alpha )}e_{\nu }^{(\beta )}\eta _{(\alpha )(\beta
)}.  \label{10}
\end{equation}
The covariant derivative in Eq.(\ref{8}) is 
\begin{equation}
\nabla _{\mu }=\partial _{\mu }+\Sigma _{\mu },  \label{11}
\end{equation}
with spinorial connections which can be written as 
\begin{equation}
\Sigma _{\mu }=\Gamma _{\mu }\otimes \text{I}+\text{I}\otimes \Gamma _{\mu },
\label{12}
\end{equation}
where 
\begin{equation}
\Gamma _{\rho }=\frac{1}{8}\left[ \widetilde{\gamma }^{(i)},\widetilde{%
\gamma }^{(k)}\right] e_{(i)}^{\nu }e_{(k)\nu ;\rho }.  \label{13}
\end{equation}
In the massless limit of DKP equation the particle-antiparticle are
identical and hence mass eigenvalue is zero. Therefore, DKP equation reduces
to $4\times 4$ massless DKP equation as follow \cite{20} 
\begin{equation}
\beta ^{\mu }\nabla _{\mu }\Psi =0,  \label{14}
\end{equation}
where $\beta ^{\mu }$ are now 
\begin{equation}
\beta ^{\mu }(x)=\sigma ^{\mu }(x)\otimes \text{I}+\text{I}\otimes \sigma
^{\mu }(x)  \label{15}
\end{equation}
with $\sigma ^{\mu }(x)=($I$,\overrightarrow{\sigma }(x))$ and 
\begin{equation}
\nabla _{\mu }=\partial _{\mu }+\Sigma _{\mu },  \label{16}
\end{equation}
where spinorial connections $\Sigma _{\mu }$ are given with the limit $%
\gamma ^{\mu }\rightarrow \sigma ^{\mu }$ as

\begin{equation}
\Sigma _{\mu }=%
\mathrel{\mathop{\lim }\limits_{\gamma \rightarrow \sigma }}%
\Gamma _{\mu }\otimes \text{I}+\text{I}\otimes \Gamma _{\mu }.  \label{17}
\end{equation}

In this paper we study solution of the photon equation (\ref{14}) in a
singular homogeneous stationary G\"{o}del-type and a stationary G\"{o}del
universes. In Section II we write down four coupled equations for the
general stationary G\"{o}del-type universes and then solved for singular
homogeneous G\"{o}del-type universe. In Section III the equations obtained
in Section II is considered for G\"{o}del universe and solved. In Section IV
both results are discussed and quantum mechanical oscillatory regions are
found. Finally energy spectrums are obtained.

\section{Solution of Photon Equation for the G\"{o}del-type Universe}

For the line-element given Eq.(\ref{2}) it must be introduced tetrads with a
suitable selection. For the simplicity we can choose 
\begin{equation}
e_{(0)}^{\mu }=\delta _{0}^{\mu },e_{(1)}^{\mu }=\delta _{1}^{\mu
},e_{(3)}^{\mu }=\delta _{3}^{\mu },e_{(2)}^{\mu }=\frac{1}{D}(\delta
_{2}^{\mu }-m\delta _{0}^{\mu })  \label{18}
\end{equation}
where $x^{0}=t,x^{1}=r,x^{2}=\theta ,x^{3}=z\cite{17}.$

The curved Dirac matrices which satisfy 
\begin{equation}
\{\gamma ^{\mu },\gamma ^{\nu }\}=2g^{\mu \nu }  \label{19}
\end{equation}
are given by $(\gamma ^{\mu }(x)=e_{(\alpha )}^{\mu }\widetilde{\gamma }%
^{(\alpha )})$%
\begin{equation}
\gamma ^{0}=\widetilde{\gamma }^{0}-\frac{m}{D}\widetilde{\gamma }%
^{2},\gamma ^{1}=\widetilde{\gamma }^{1},\gamma ^{2}=\frac{1}{D}\widetilde{%
\gamma }^{2},\gamma ^{3}=\widetilde{\gamma }^{3}.  \label{20}
\end{equation}
The spinorial connections are 
\begin{equation}
\Gamma _{0}=\frac{m^{\prime }}{4D}\widetilde{\gamma }^{2}\widetilde{\gamma }%
^{1},\Gamma _{1}=\frac{m^{\prime }}{4D}\widetilde{\gamma }^{0}\widetilde{%
\gamma }^{2},\Gamma _{2}=-\frac{m^{\prime }}{4}\widetilde{\gamma }^{1}%
\widetilde{\gamma }^{0}+\frac{mm^{\prime }+l^{\prime }}{4D},\Gamma _{3}=0
\label{21}
\end{equation}
where prime indicates derivative with respect to $r.$ Using Eq.(\ref{14})
and Eq.(\ref{15}) and Jauch and Rohrlich\cite{23} representation of Dirac
matrices we obtain the photon equation as 
\[
\left\{ \left[ -2i(\text{I}\otimes \text{I})-\frac{m}{D}(\widetilde{\sigma }%
^{2}\otimes \text{I}+\text{I}\otimes \widetilde{\sigma }^{2})\right]
\partial _{t}+(\widetilde{\sigma }^{1}\otimes \text{I}+\text{I}\otimes 
\widetilde{\sigma }^{1})\partial _{r}+\frac{1}{D}(\widetilde{\sigma }%
^{2}\otimes \text{I}+\text{I}\otimes \widetilde{\sigma }^{2})\partial
_{\theta }\right. 
\]
\[
+(\widetilde{\sigma }^{3}\otimes \text{I}+\text{I}\otimes \widetilde{\sigma }%
^{3})\left( \partial _{z}-\frac{m^{\prime }}{2D}\right) +\frac{iD^{\prime }}{%
2D}(\widetilde{\sigma }^{2}\otimes \text{I}+\text{I}\otimes \widetilde{%
\sigma }^{2})(\widetilde{\sigma }^{3}\otimes \text{I}+\text{I}\otimes 
\widetilde{\sigma }^{3}) 
\]
\begin{equation}
\left. -\frac{im^{\prime }}{4D}\left[ (\widetilde{\sigma }^{1}\otimes \text{I%
}+\text{I}\otimes \widetilde{\sigma }^{1})(\widetilde{\sigma }^{2}\otimes 
\text{I}+\text{I}\otimes \widetilde{\sigma }^{2})-(\widetilde{\sigma }%
^{2}\otimes \text{I}+\text{I}\otimes \widetilde{\sigma }^{2})(\widetilde{%
\sigma }^{1}\otimes \text{I}+\text{I}\otimes \widetilde{\sigma }^{1})\right]
\right\} \Psi =0.  \label{22}
\end{equation}
To obtain stationary solutions it is used a function as follows 
\begin{equation}
\Psi =\exp \left[ -i(\omega t+k_{2}\theta +k_{3}z)\right] \Phi ,  \label{23}
\end{equation}
hence Eq.(\ref{22}) gives four coupled differential equations in terms of
the components of the spinor as 
\begin{equation}
\partial _{r}(\Phi _{B}+\Phi _{C})+\frac{1}{D}(m\omega -k_{2})(\Phi
_{B}+\Phi _{C})-2(\omega +ik_{3})\Phi _{A}=0,  \label{24}
\end{equation}
\begin{equation}
\left( \partial _{r}-\frac{D^{\prime }}{D}\right) (\Phi _{A}+\Phi _{D})-%
\frac{1}{D}(m\omega -k_{2})(\Phi _{A}-\Phi _{D})-2\omega \Phi _{B}=0,
\label{25}
\end{equation}
\begin{equation}
\left( \partial _{r}-\frac{D^{\prime }}{D}\right) (\Phi _{A}+\Phi _{D})-%
\frac{1}{D}(m\omega -k_{2})(\Phi _{A}-\Phi _{D})-2\omega \Phi _{C}=0,
\label{26}
\end{equation}
\begin{equation}
\partial _{r}(\Phi _{B}+\Phi _{C})-\frac{1}{D}(m\omega -k_{2})(\Phi
_{B}+\Phi _{C})-2(\omega -ik_{3})\Phi _{A}=0.  \label{27}
\end{equation}
From Eq.(\ref{25}) and Eq.(\ref{26}) $\Phi _{B}=\Phi _{C}.$ Up to now we
have not used an explicit form of $m(r)$ and $l(r)$. Hence, different
choices of $m$ and $l$ lead us to different rotating models.

For the Eq.(\ref{5}) we have that $m=m_{0}+CAr^{2}/2$ and this is the
singular homogeneous G\"{o}del-type universe and Eqs.(\ref{24})-(\ref{27})
reduces to 
\begin{equation}
\lbrack \partial _{r}+\frac{1}{r}(\widetilde{\omega }+\beta r^{2})]\Phi _{B}-%
\frac{1}{\alpha }\Phi _{A}=0,  \label{28}
\end{equation}
\begin{equation}
\left( \partial _{r}-\frac{1}{r}\right) (\Phi _{A}+\Phi _{D})-\frac{1}{r}(%
\widetilde{\omega }+\beta r^{2})(\Phi _{A}-\Phi _{D}),  \label{29}
\end{equation}
\begin{equation}
\lbrack \partial _{r}-\frac{1}{r}(\widetilde{\omega }+\beta r^{2})]\Phi _{B}-%
\frac{1}{\alpha ^{\ast }}\Phi _{D}=0,  \label{30}
\end{equation}
where 
\begin{equation}
\widetilde{\omega }=\frac{m_{0}\omega }{A}-\frac{k_{2}}{A},\beta =\frac{C}{2}%
,\alpha =(\omega +ik_{3})^{-1}.  \label{31}
\end{equation}
Eliminating $\Phi _{A}$ and $\Phi _{D}$ from Eq.(\ref{28}) and Eq.(\ref{30})
we obtain following second order differential equation for $\Phi _{B}$ as 
\begin{equation}
\partial _{r}^{2}\Phi _{B}-\frac{1}{r}\partial _{r}\Phi _{B}+\left( -\frac{H%
}{r^{2}}-E-\beta ^{2}r^{2}\right) \Phi _{B}=0,  \label{32}
\end{equation}
where 
\begin{equation}
E=\widetilde{\omega }C+\omega ^{2}+k_{3}^{2},H=\widetilde{\omega }^{2}-\frac{%
2\sqrt{2}ik_{3}\widetilde{\omega }}{\omega }.  \label{33}
\end{equation}
Introducing a variable as $u=\beta r^{2}$ it is found that 
\begin{equation}
4u^{2}\partial _{u}^{2}\Phi _{B}+\left( -u^{2}-H-\frac{E}{\beta }u\right)
\Phi _{B}=0,  \label{34}
\end{equation}
where if it is used for constants $H=4\lambda ^{2}-1$ and $4\kappa =-E/\beta
,$ one can be obtained well-known Whittaker differential equation has
solution \cite{24} 
\begin{equation}
\Phi _{B}(u)=B_{1}W_{\kappa ,\lambda }(u)+B_{2}M_{\kappa ,\lambda }(u).
\label{35}
\end{equation}

\section{Solution of Photon Equation for the G\"{o}del Universe}

As was mentioned before the G\"{o}del space-time can be obtained from the
G\"{o}del-type space-time by taking $m=\exp (ar)$ and $l=(-1/2)\exp (2ar)$
in Eq.(\ref{2}). So it is sufficient to introduce these values of $m$ and \ $%
l$ in Eqs.(\ref{24})-(\ref{27}) to obtain solution of photon equation for
the G\"{o}del space-time. Then we obtain following coupled equations 
\begin{equation}
\partial _{r}\Phi _{B}+\sqrt{2}(\omega -k_{2}e^{-ar})\Phi _{B}-\frac{1}{%
\alpha }\Phi _{A}=0,  \label{36}
\end{equation}
\begin{equation}
(\partial _{r}-a)(\Phi _{A}+\Phi _{D})-\sqrt{2}(\omega -k_{2}e^{-ar})(\Phi
_{A}-\Phi _{D})-2\omega \Phi _{B}=0,  \label{37}
\end{equation}
\begin{equation}
\partial _{r}\Phi _{B}-\sqrt{2}(\omega -k_{2}e^{-ar})\Phi _{B}-\frac{1}{%
\alpha ^{\ast }}\Phi _{D}=0,  \label{38}
\end{equation}
where again $\Phi _{B}=\Phi _{C}.$ Substituting Eq.(\ref{36}) and Eq.(\ref
{38}) in Eq.(\ref{37}) it is found second order differential equation as
follows 
\begin{equation}
\partial _{r}^{2}\Phi _{B}+\left( -2k_{2}^{2}e^{-2ar}+Ae^{-ar}-B\right) \Phi
_{B}=0,  \label{39}
\end{equation}
where 
\begin{equation}
A=4\omega -\frac{i2\sqrt{2}ak_{3}}{\omega },B=3\omega ^{2}+k_{3}^{2}-i\sqrt{2%
}k_{3}a.  \label{40}
\end{equation}
If variable is changed as $u=\frac{\sqrt{8}k_{2}e^{-ar}}{a}$ and a function$%
\ \Phi _{B}=u^{-1}g_{B}$ is used, the following Whittaker differential
equation is obtained 
\begin{equation}
4u^{2}\partial _{u}^{2}g_{B}+[-u^{2}+4\sigma u-(4\rho ^{2}-1)]g_{B}=0.
\label{41}
\end{equation}
The solution of the Eq.(\ref{41}) can be written as 
\begin{equation}
g_{B}(u)=C_{1}W_{\sigma ,\rho }(u)+C_{2}M_{\sigma ,\rho }(u)  \label{42}
\end{equation}
with 
\begin{equation}
4\sigma =\frac{\sqrt{2}A}{a}\,,\rho ^{2}=\frac{B}{4a^{2}}+\frac{1}{4}.
\label{43}
\end{equation}

\section{Exact Solutions}

The solutions Eq.(\ref{35})and Eq.(\ref{42}) can be written in terms of the
confluent hypergeometric functions $U$ and $M$ using 
\begin{equation}
W_{\alpha ,\xi }(x)=e^{-x/2}x^{\xi +1/2}U(\frac{1}{2}+\xi -\alpha ,1+2\xi
;x),  \label{44}
\end{equation}
\begin{equation}
M_{\alpha ,\xi }(u)=e^{-x/2}x^{\xi +1/2}M(\frac{1}{2}+\xi -\alpha ,1+2\xi
;x),  \label{45}
\end{equation}
so exact solution of photon equation for the singular homogeneous
G\"{o}del-type universe is 
\[
\Psi _{B}=e^{-i(\omega t+k_{2}\theta +k_{3}z)}e^{-(\frac{Cr^{2}}{4})}(\frac{%
Cr^{2}}{2})^{\lambda +1/2}[B_{1}U(\frac{1}{2}+\lambda -\kappa ,1+2\lambda ;%
\frac{Cr^{2}}{2}) 
\]
\begin{equation}
+B_{2}M(\frac{1}{2}+\lambda -\kappa ,1+2\lambda ;\frac{Cr^{2}}{2})].
\label{46}
\end{equation}
Similarly solution of the photon equation for the G\"{o}del universe is 
\[
\Psi _{B}=(\frac{\sqrt{8}k_{2}e^{-ar}}{a})^{\rho -1/2}e^{-i(\omega
t+k_{2}\theta +k_{3}z)}e^{-(\frac{\sqrt{8}k_{2}e^{-ar}}{2a})} 
\]
\begin{equation}
\times \lbrack C_{1}U(\frac{1}{2}+\rho -\sigma ,1+2\rho ;\frac{\sqrt{8}%
k_{2}e^{-ar}}{a})+C_{2}M(\frac{1}{2}+\rho -\sigma ,1+2\rho ;\frac{\sqrt{8}%
k_{2}e^{-ar}}{a})].  \label{47}
\end{equation}

From the condition on Whittaker functions that must be bounded for all
values of variable we find energy spectrum for both space-times as follows 
\begin{equation}
\frac{1}{2}+\lambda -\kappa =-n_{1,}  \label{48}
\end{equation}
\begin{equation}
\frac{1}{2}+\rho -\sigma =-n_{2}.  \label{49}
\end{equation}
where $n_{1}$ and $n_{2}$ are positive integers or zero. The explicit forms
of Eqs.(\ref{48}) and (\ref{49}) are 
\begin{equation}
n_{1}=-\frac{\omega ^{2}}{2C}-\frac{k_{3}^{2}}{2C}-\frac{\widetilde{\omega }%
}{2}\mp \sqrt{\widetilde{\omega }^{2}-2\sqrt{2}ik_{3}\frac{\widetilde{\omega 
}}{\omega }+1}-\frac{1}{2},  \label{50}
\end{equation}
\begin{equation}
n_{2}=\frac{\sqrt{2}\omega }{a}\mp \frac{1}{2a}\sqrt{3\omega
^{2}+k_{3}^{2}-ia\sqrt{2}k_{3}+a^{2}}-\frac{ik_{3}}{\omega }-\frac{1}{2}.
\label{51}
\end{equation}

One can show that solutions (\ref{46}) and (\ref{47}) are oscillatory in
behavior only regions (for detail Ref.\cite{12}) 
\[
-\frac{1}{2}(\omega ^{2}+\widetilde{\omega }C+k_{3}^{2})-\sqrt{2(\omega ^{2}+%
\widetilde{\omega }C+k_{3}^{2})-\frac{\widetilde{\omega }^{2}C}{2}+\sqrt{2}i%
\frac{k_{3}\widetilde{\omega }C}{\omega }}<u< 
\]
\begin{equation}
-\frac{1}{2}(\omega ^{2}+\widetilde{\omega }C+k_{3}^{2})+\sqrt{2(\omega ^{2}+%
\widetilde{\omega }C+k_{3}^{2})-\frac{\widetilde{\omega }^{2}C}{2}+\sqrt{2}i%
\frac{k_{3}\widetilde{\omega }C}{\omega }},  \label{52}
\end{equation}
\[
\omega -\frac{iak_{3}}{\sqrt{2}\omega }-\sqrt{(\omega -\frac{iak_{3}}{\sqrt{2%
}\omega })^{2}-\frac{3\omega ^{2}}{2}-\frac{k_{3}^{2}}{2}+\frac{iak_{3}}{%
\sqrt{2}}}<k_{2}u< 
\]
\begin{equation}
\omega -\frac{iak_{3}}{\sqrt{2}\omega }+\sqrt{(\omega -\frac{iak_{3}}{\sqrt{2%
}\omega })^{2}-\frac{3\omega ^{2}}{2}-\frac{k_{3}^{2}}{2}+\frac{iak_{3}}{%
\sqrt{2}}.}  \label{53}
\end{equation}

\section{\protect\bigskip Results and Discussions}

We have analyzed the photon equation in the background of stationary
G\"{o}del-type and the G\"{o}del universes. The method of separation of
variables is used because of the simple symmetry of the G\"{o}del universes.
The energy spectrum of particle and oscillatory character of solutions were
found.

The results obtained can be used to study quantum field theory in curved
rotating space-times. Also it is necessary to quantize obtained wave
functions to discuss pair creation and annihilation. In addition, these
results can be compared with earlier results of electromagnetic fields in
G\"{o}del space-times and calculate quantum corrections on these results.
The effects of rotation on propagation and helicity of electromagnetic
fields can be obtained by using the wave functions found in the Section II
and Section III.

\end{document}